# Evaporation of initially heated sessile droplets and the resultant dried colloidal deposits on substrates held at ambient temperature

Sanghamitro Chatterjee, Manish Kumar, Janani Srree Murallidharan and Rajneesh Bhardwaj[*]
Department of Mechanical Engineering, Indian Institute of Technology Bombay, Mumbai 400076, India
*Corresponding author (rajneesh.bhardwaj@iitb.ac.in, Phone: +91 22 2576 7534)

## Abstract

The present study experimentally and numerically investigates the evaporation and resultant patterns of dried deposits of aqueous colloidal sessile droplets, when the droplets are initially elevated to a high temperature before being placed on a substrate held at ambient temperature. The system is then released for natural evaporation without applying any external perturbation. Infrared thermography and optical profilometry were used as essential tools for interfacial temperature measurements and quantification of the coffee-ring dimensions, respectively. Initially, a significant temperature gradient exists along the liquid-gas interface as soon as the droplet is deposited on the substrate which triggers a Marangoni stress-induced recirculation flow directed from the top of the droplet towards the contact line along the liquid-gas interface. Thus, the flow is in the reverse direction to that seen in the conventional substrate heating case. Interestingly, this temperature gradient decays rapidly- within the first 10% of the total evaporation time and the droplet-substrate system reaches thermal equilibrium with ambient thereafter. Despite fast decay of the temperature gradient, the coffee-ring dimensions significantly diminish, leading to an inner deposit. A reduction of 50-70% in the coffee-ring dimensions was recorded by elevating the initial droplet temperature from 25ºC to 75ºC, for suspended particle concentration varying between 0.05% v/v to 1.0% v/v. This suppression of the coffee-ring effect is attributed to the fact that the initial Marangoni stress-induced recirculation flow continues until the last stage of the evaporation, even after the interfacial temperature gradient vanishes. This is essentially a consequence of liquid inertia. Finally, a finite-element based two-dimensional modeling in axisymmetric geometry has been found to capture the measurements with reasonable fidelity and the hypothesis considered in the present study corroborates well with a first approximation qualitative scaling analysis. Overall, together with a new experimental condition, the present investigation discloses a distinct nature of Marangoni stress-induced flow in the drying droplet and its role in influencing the associated colloidal deposits, which was not explored previously. The insights gained from this study are useful to advance technical applications such as spray cooling, ink-jet printing, bioassays, etc.

## Introduction

The seemingly simple problem of evaporation of a sessile water droplet containing suspended colloidal particles has attracted significant attention over the last decade. This is evident from the importance attributed to the conventional "coffee ring" effect (CRE), in which a ring-like solid stain is left behind the evaporating droplet [1]. Formation of such a ring-like pattern was first demonstrated analytically and experimentally based on a fluid flow theory by drawing an analogy to classical electrostatic problem [1-2]. Since then, the phenomenon has been the subject of research in both applied and fundamental sciences. Depending upon the particular application, CRE may either be preferential as well as detrimental [3]. CRE is typically utilized in self-ordered ring based processes that require an increase in analyte concentration along the droplet periphery for several applications in biomedical, forensics, and diagnostics [4]. Besides, several industrial applications such as patterning of particles [5] for the fabrication of electronic components [6] require fine control over the CRE self-assembly process. In contrast, applications such as inkjet printing, surface coating, coolant deposition [7, 8, 9] require suppression of the coffee ring effect. On the other hand, fundamental research still demands to understand the basic mechanisms and rules that govern the evaporation dynamics under different conditions [10, 11, 12]. Therefore, several efforts have been devoted to understand and control the evaporation mechanism of sessile droplets for various technical applications [13].

The evaporation dynamics of a sessile droplet depends upon the motion of the triple-phase contact line and, in general, two modes of motions are found [3], the constant-contact-radius (CCR) mode, wherein, the contact line of the droplet is pinned, and the constant-contact-angle (CCA) mode stems from depinning of the triple-phase contact line. Also, a "stick-slip" motion [12] is also observed, whose molecular origin has been demonstrated [14]. Substrate hydrophobicity is learned to play an important role in deciding the mode of evaporation and the overall lifetime of the evaporating droplet [15]. The flow mechanisms at different conditions in an evaporating droplet execute several interesting features, namely, evaporation-triggered segregation of sessile binary droplets [16], CRE driven hollow rims formation at the periphery of evaporating pure water drop on salt substrates [17], suppression of vaporization rate and the emergence of asymptotic universal pattern for two adjacent droplets [18] and interesting phenomena observed in inclined droplets [19]. Besides, vapor bubble formation for evaporating droplets resting on superheated surfaces having varied wettability was demonstrated [20]. Besides, pinning free droplet evaporation [21], and the effect of surface roughness and droplet size on the transition between different evaporation modes have been looked at [22].

Thus, the evaporation of sessile droplet and subsequent stain formation involves complex mechanism [23] and the underlying physics is largely exploited to control CRE. The ways of controlling CRE fall into the following categories: Contact line depinning by introducing surface texture and hydrophobicity [24], AC electric field-induced electrowetting[25, 26], wherein, the periodic motion of the contact line under the electric field and generation of internal flow at higher frequencies result in suppression of CRE. However, the application of the DC electric field was proven to induce smoother receding of the contact line of nanofluid droplets and more uniform deposits owing to a different mechanism of nanoparticle interaction with the applied electric potential [27]. Tuning the evaporation flux distribution by geometrically controlling the droplet shape and the subsequent preferential deposition pattern have been demonstrated as a way to tailor CRE [28]. Evaporation in a controlled environment and the consequent surface capturing effect has also been proven to be a viable tool for suppressing CRE [29]. Apart from these, Marangoni convection induced modification in evaporation dynamics, and thereby modifying the resultant deposition pattern is a promising tool for suppression of CRE. Marangoni flow can be generated by introducing surfactants [30] or by using saline solution [31]. However, these are detrimental where the introduction of additives to the liquid is not desired. Yen *et. al*. introduced a distinct method for suppressing the CRE by reversing the flow direction brought about by differential evaporation generated by a focused laser beam at the droplet apex point [32] which avoids introducing additives to the drop and also invading substrate surface morphology. Besides, thermal Marangoni flow generation and the resultant deposition pattern have been explored by substrate heating wherein, a transition from "coffee ring" to "coffee eye" and a diminished outer ring was obtained[33, 34].

As it appears, several efforts have been devoted to generating thermal Marangoni-induced circulation flow mostly by heating the substrate. To the best of authors' knowledge, there is no literature available wherein, the elevation of droplet temperature, while keeping the substrate temperature as that of ambient, and subsequent temperature gradient induced flow physics have been looked at. In ref. [32], focusing of a high power laser beam at the droplet apex led to a localized heating and a diverging evaporative flux at the droplet apex, leading to a reversal of the coffee ring effect as well as depinning of the triple contact line. No temperature gradient induced Marangoni recirculation was reported. The present research looks at the flow physics and the deposition pattern upon evaporation of droplets containing suspended colloidal polystyrene microparticles ( ~ 1.1 µm mean diameter), given the condition that the droplets are elevated to high initial temperatures before being placed on a substrate (glass in the present case) kept at room temperature. For this purpose, the aqueous colloidal solution was first heated to achieve the desired temperature and then the heated droplets were placed on a glass substrate kept at ambient (cf. figure 1). The system is then released for natural evaporation without applying any external perturbation. Under such conditions, it is envisioned that at the early stages of evaporation, a significant temperature gradient would exist along the surface followed by a thermal equilibration with the substrate and the ambient by thermal conduction and evaporative cooling. The resultant flow physics and deposition pattern under such circumstances is unknown and is the subject of investigation of in the present work. Based on the motivation outlined above, the following research questions have been considered in the present work. (1) How will the energy and mass transport happen in such an unusual case, where an initially heated colloidal particle-laden droplet is placed on a substrate held at ambient air? (2) What will be the evaporation mechanism, and the temporal behavior of the interfacial temperature and vapor concentration field, and the general physical principles that govern them? (3) What is the net effect of the initial droplet heating on the internal microfluidic flow inside the evaporating sessile droplet and what will be the temporal behavior of the modified flow? (4) Most importantly, what will be the pattern of resultant dried deposit given this new experimental condition?

The distinction of the present work is twofold- first, in the present case, the temperature gradient along the liquid-vapor interface would be in the opposite direction and a consequent flow circulation be in the opposite direction than that of the heated substrate case[33, 34, 35]. Secondly and importantly, the present infra-red thermography study shows that the initial temperature gradient along the liquid-vapor interface decays extremely fast and the typical time scale of this decay is much less than the total evaporation time (cf. figures 7-9). Despite this fact, it was found that the coffee-ring effect was significantly suppressed and a modified deposition pattern with inner deposits was obtained (cf. figures 10-12). It is hypothesized that during early stages of evaporation, a strong temperature gradient exists along the liquid-vapor interface, giving rise to a Marangoni stress which generates a flow circulation from the droplet apex towards the contact line, resulting in an axisymmetric vortex-like flow pattern inside the drying droplet, and this recirculating flow continues till the last stage of evaporation process due to liquid inertia and is subject to viscous diffusion, thereby bringing about a significant modification in the dried deposit pattern. This hypothesis is further evidenced by optical visualization wherein, the motion of suspended fluorescent microparticles is tracked during the whole evaporation process (cf. figure 14). Besides, a model involving two-way coupling of the energy equation in a sessile droplet and substrate and the transport of liquid vapor outside the droplet for diffusion-limited, steady-state evaporation [36] has been extended here to include transient terms in the governing equations for temperature and vapor concentration field. This allowed capturing the highly transient behavior of temperature and vapor concentration field observed at the initial stage of evaporation (cf. figures 8,9). This model provides insights into the underlying mechanism that governs energy and mass transport when an initially heated sessile droplet is placed on a substrate kept at ambient temperature. Furthermore, a first approximation treatment involving vortex viscous diffusion captures the postulate that even after the temperature gradient along the liquid-vapor interface vanishes, the recirculating internal fluid flow continues till the last stage of the evaporation with viscous decay as a consequence of liquid inertia bringing about a suppression in coffee ring and formation of associated inner deposits. The findings reported in this article may be proven to be useful in technical applications where colloidal suspension solution heating may be preferred over substrate heating, e.g., substrates like aluminum that tends to oxidize upon heating in ambient, or plastic/soft materials that melt or deform at elevated temperatures or in processes where multiple droplets are involved and control over individual droplets is required, e.g., evaporation of droplets of matrix/analyte solution in MALDI-TOF mass spectroscopy.

## Methods

### Experimental details

*Preparation of colloidal suspensions:*

An aqueous colloid suspension with a concentration of 10% v/v of uniformly dispersed polystyrene (PS) latex microparticles having a mean diameter of ~ 1.1 µm was procured from Sigma-Aldrich Corporation, USA (product: LB11). The properties of the particles can be found in the manufacturer product information sheet [37]. The zeta potential of the particles at a concentration of 0.05% v/v in aqueous solution (pH 6.5-7, conductivity ~ 0.25 mS/cm) was measured at 25°C by Laser Doppler Electrophoresis method [38] (Zetasizer Nano ZS, Malvern Panalytical, UK) by using a cuvette equipped with gold platted electrode (DTS1070) and was found to be ~ $-25.5 \pm 1.5$ mV which is consistent with previous measurements [39]. This solution was further diluted by deionized (DI) water to achieve solutions of three different particle concentration $c' = 0.05\%, 0.1\%$ and $1.0\%$, which were used as working concentrations. The solutions were sufficiently stabilized and it was ensured that no sedimentation and agglomeration of particles would occur during storage in microcentrifuge tubes (Eppendorf. Inc.).

To track the internal fluid flow during evaporation, fluorescent polystyrene microparticles (mean diameter ~ 1.0 µm) suspended colloidal solutions having a particle concentration of 1 % v/v was obtained from Fisher Scientific, USA (product B0100), which were further diluted to a particle concentration $c'$ of 0.1 % v/v. Necessary information about the product is available in the manufacturer product information sheet [40]. The zeta potential of the particles was found to be $-38.2 \pm 2$ mV. The particles are fluorescent blue having excitation and emission wavelengths of 365 nm and 447 nm respectively, which is compatible with the fluorescence filter used (Olympus, U-FUNA, excitation filter: BP360-370, emission filter: BA420-460 and dichromatic mirror DM410) for observation.

*Elevation of initial droplet temperature ($T_i$):*

Figure 1 shows the schematic diagram for heating of colloidal solution and dispensing of a droplet having initial elevated temperature on a glass substrate kept at ambient conditions. A glass centrifuge tube containing the aqueous colloidal solution was immersed in a pool of normal water kept at a constant high temperature within a temperature bath (Cole-Parmer StableTemp, model- WB02). The temperature of the solution was measured by a thermocouple (TC direct, 818-044). The temperature data was acquired by NI-DAQ (National Instruments) data acquisition system (NI 9212, 8 channel C series temperature input module) integrated with Labview software. After the desired temperature is reached, droplets of the heated liquid were dispensed on glass substrates with the help of a micropipette (PRIME, Biosystem diagnostics Inc., India, ISO 9001 2008). The micropipette tip was immersed in the solution for a sufficiently long time before withdrawing the solution, to ensure that the solution inside the tip remains at the desired temperature. The initial high temperature of the dispensed droplet was further observed by Infra-red (IR) camera, cf. figure 2. It was found that the experimental difference between thermocouple temperature measurements and IR thermography is not more than $\pm\,4-5°C$ (cf. figures 7, 8, and supplementary movies V2-V5), thus ensuring proper experimental conditions, as desired.

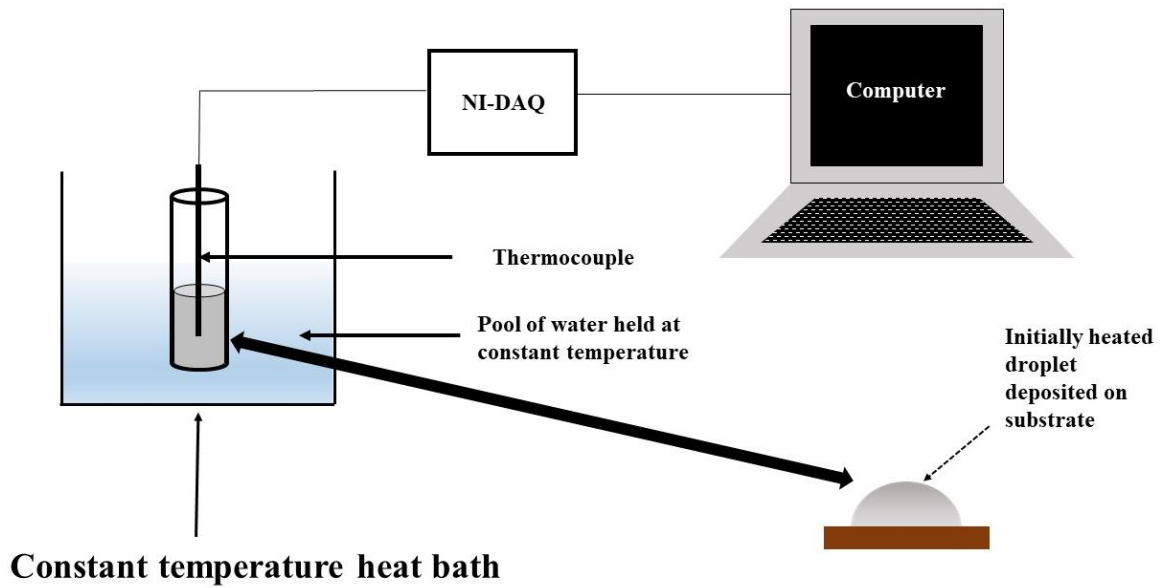

**Figure 1.** The schematic diagram for colloidal suspension solution heating and temperature measurements

*High-speed visualization and infrared thermography:*

Figure 2 shows the experimental setup. An initially heated droplet of volume $1.1 \pm 0.2\ \mu L$ was gently dispensed on glass slides of dimensions $75 \times 15 \times 1.2\ mm^3$ (Borosil India Ltd., 9100P02, ISO 8037) kept at ambient conditions. As shown in figure 2, the droplet is illuminated by an LED light source from one side and the temporal geometry of the droplet is recorded from the opposite direction by using a high-speed camera (IDT Inc, MotionPro Y3 classic) at a frame rate of 10 frames/second (FPS). The working distance, image resolution, and pixel resolution were kept 9.5 cm, 600 × 400, and 14 μm respectively. The data of the recorded images in terms of contact angle and wetted diameter were analyzed using MATLAB image processing module [34, 41]. The reflection of the droplet onto the substrate is visible, providing the advantage to detect the contact points accurately. On each droplet image, the contact angle values were obtained at the contact points both on the left and right sides of the droplet. The contact angle measurements at these two points exhibit a nominal difference (~2°) and hence, an average of these two measurements have been taken. This way, measurements have been performed on randomly chosen different locations on the substrate surface to check the reproducibility of the measurements as well as to ensure the uniformity of the substrate surface. The experimental spread has been shown by error bars (cf. figures 4, 6). Spatial as well as temporal temperature distribution was recorded from the top by a vertically mounted IR camera (A670sc, FLIR systems Inc.) with its IR lens (25 mm, f/2.5) looking down (cf. figure 2). The imaging rate was kept at 100 FPS, as the temperature field exhibits a highly transient behavior at the initial stages of evaporation (cf. figures 7, 8). Hence, this method enables us to measure both the liquid-vapor and solid-vapor interfacial temperatures. The working distance, image resolution, and pixel resolutions are 10.6 cm, 300×256, and 16 μm respectively. The uncertainty in the temperature measurements is ±0.6°C, and the emissivity of water and glass being 0.97 and 0.95 respectively. The calibration of IR temperature response by taking into account the above emissivity values can be found elsewhere [34]. During experiments, the ambient temperature and relative humidity were kept fixed at 25±2°C and 40±5% respectively.

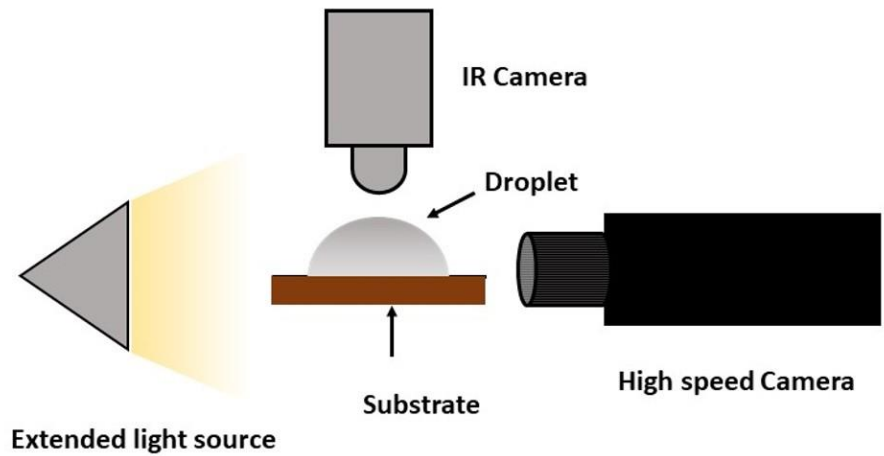

**Figure 2.** Schematic of the experimental setup showing different components

*Preparation of substrate:*

The glass substrates were successively cleaned by acetone, isopropanol, and deionized water and were dried by blowing dry nitrogen. The surface quality was further examined by Atomic Force Microscopy (AFM) (MFP:3D classic AFM (Asylum Research)). The AFM tip material is n-type Silicon (resistivity=0.01–0.025 Ohm-cm) with aluminum as detector coating and has a nominal radius of curvature ~9±2 nm. AFM scans of scan area 1 μm$^2$ were performed in ambient conditions in tapping mode at different locations on the samples to account for all types of irregularities and the average values of Root Mean Square (RMS) roughness have been reported along with the experimental spread.

Figure 3 shows representative AFM measurements of the prepared glass surfaces. The RMS roughness and the surface area factor ($r$) were found to be 0.61± 0.06 nm and 1.002-1.005 respectively. Hence, the surface exhibits nominal roughness and is of good quality. Further, contact angle hysteresis was measured on these surfaces by the tilting plate method. Under ambient conditions, the apparent contact angle was 30˚±2˚ and the advancing and the receding contact angles were found to be ~ 40˚±3˚ and 15˚±5˚ respectively. Such order of hysteresis is desired on surfaces having nominal roughness [42], as in the present case. Therefore, the pinning of the triple-phase contact line and subsequent formation coffee stain pattern is expected upon the evaporation of colloidal particle-laden sessile droplets on the surfaces.

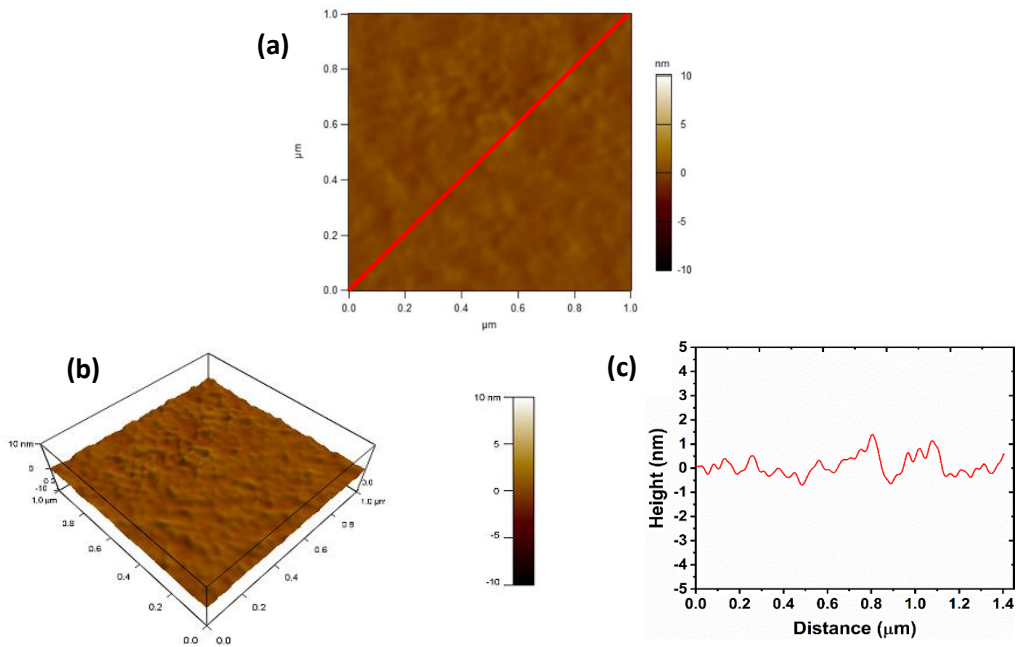

**Figure 3.** Representative AFM measurements of the prepared glass substrate (a) AFM image, (b) perspective view, and (c) surface line profile along diagonal shown in (a)

*Visualization of deposits, particle tracking, and measurement of deposit dimensions:*

For visualization of stain deposits, the aforesaid high-speed camera was mounted on an optical microscope (Olympus MX-40, objective lens 4X, and 40X). The microscope is also equipped with fluorescence filter (Olympus, U-FUNA, excitation filter: "BP360-370", emission filter: "BA420-460" and dichromatic mirror "DM410") for observation of the motion of fluorescent particles suspended in the droplet and their motion were captured at 2 FPS using an objective lens 10X. The ring profiles were obtained by using a 3D optical profilometer (Zeta-20, Zeta Instruments Inc.) at four azimuthal locations and were averaged. The experimental spread has been presented by error bars (cf. figure 13).

## Computational Model

A computational model has been developed to account for the time-varying temperature and liquid-vapor concentration field in axisymmetric cylindrical coordinates. A model involving two-way coupling of the energy equation in a sessile droplet and substrate, and the transport of liquid vapor outside the droplet for diffusion-limited, steady-state evaporation [36], has been extended here to include the transient term in the governing equations. The two-way coupling method couples the energy transport in the droplet-substrate and evaporative cooling due to vapor diffusion in the ambient [36, 43]. Inclusion of the transient terms is important to capture the highly transient behavior of the temperature field during the initial stage of the evaporation (cf. figures 7-9). The diffusion-limited evaporation is justified here, as it was conclusively proven earlier that the diffusion governs the evaporation, regardless of the complexity of mechanism [44].

We neglect advection inside the droplet and this is justified for low Péclet number (*Pe*). We estimate *Pe* based on the Marangoni flow velocity ($v_{ma}$) obtained in the measurements of the droplets considered in the present study (O(10$^{-4}$ m/s)). *Pe* is ~ 0.7 based on $v_{ma}$ and as pointed out by Larson [45] that heat transfer due to advection is significant if *Pe* exceeds 10. The energy (or temperature *T*) equation for the droplet (*j = d*) and the substrate (*j = s*) is therefore given by,

$$\rho_j c_{p,j} \frac{\partial T_j}{\partial t} = k_j \nabla^2 T_j, \quad (1)$$

where, $\rho$, $c_p$ and $k$ are the density, specific heat at constant pressure, and thermal conductivity, respectively. Perfect thermal contact between the droplet and the substrate surface is assumed. Along the bottom of the substrate, a constant temperature boundary condition ($T = T_s$) is applied and jump energy boundary condition is applied along the liquid-vapor interface, which reads as follows,

$$jL = -k_d \nabla T . \hat{n}, \quad (2)$$

where $j$ is evaporative mass flux (kg/m$^2$-s) at the liquid-vapor interface, $L$ is the latent heat of evaporation (J/kg) and $\hat{n}$ is the unit normal on the liquid-vapor interface. In the absence of external convection, the liquid-vapor concentration ($c$, in kg/m$^3$) in surrounding gas is governed by the following diffusion transport equation,

$$\frac{\partial c}{\partial t} = D \nabla^2 c, \quad (3)$$

where $D$ is the diffusion coefficient of liquid-vapor in gas at surrounding ambient temperature. The boundary conditions and the computational domain used to solve Eqs. 1 and 3 and thermophysical quantities of the substrate (glass) and liquid (water) are described elsewhere [36].

The evaporative mass flux $j$ at the liquid-vapor interface is expressed as follows,

$$j(r,t) = -D(T)[\nabla c . \hat{n}]_{LG}, \quad (4)$$

where $D(T)$ is the diffusion coefficient of liquid-vapor in gas as a function of liquid-vapor interface temperature (m$^2$/s). The diffusion coefficient, $D(T)$, of the water vapor in ambient air is expressed as a function of temperature:

$$D(T) = 2.5 \times 10^{-4} exp\left(-\frac{684.15}{T+273.15}\right), \quad (5)$$

and the water vapor saturation concentration $c_{sat}(T)$ as a function of temperature $T$ is expressed as follows,

$$c_{sat}(T) = [9.99 \times 10^{-7} T^3 - 6.94 \times 10^{-5} T^2 + 3.20 \times 10^{-3} T - 2.87 \times 10^{-2}], \quad (6)$$

We employed the Galerkin finite-element method using MATLAB© to solve the governing equations (Eqs. 1-3) to obtain temperature ($T$) and the vapor concentration field in droplet-substrate and in ambient respectively. The algorithm in a time-step is briefly described as follows:

1) Solve two-way coupled energy and liquid-vapor transport equation with jump energy boundary condition (Eq. 2). The coupling scheme used here is described elsewhere [36].
2) Based on the evaporation flux obtained using equation 4 in step 1, calculate the evaporation rate and the remaining droplet volume. Compute the change in the droplet height, assuming it as a spherical cap. Generate a revised mesh of the droplet and ambient. Map the numerical solution from the old to the new mesh.

The mesh dimensions used in the present study are given elsewhere [36] and the mesh size independence test can be found therein. We continue the computations until the droplet volume is smaller than a threshold value (contact angle ~ 3˚). To set up the simulation, we consider diffusion-limited, transient evaporation of a sessile droplet initially at elevated temperature with a pinned contact line on a hydrophilic substrate. The droplet geometry parameters at $t = 0$ have been taken from experimental data.

## Results and Discussions

In this section, the results for the evaporation and the resultant colloidal stain pattern formed after the evaporation of colloidal sessile deionized water droplets will be presented. First, the results for colloidal particle size 1.1 µm and concentration $c' = 0.1\%$ v/v will be presented. Thereafter, results for two other concentrations: 0.05% v/v and 1.0% v/v will be presented. Thereafter, tracking of fluorescent particles for understanding the internal fluid flow will be presented, which is followed by the results of the computational model and theoretical arguments.

### Evaporation and resultant deposition pattern for concentration $c' = 0.1\%$ v/v

#### *(i) Temporal variation of droplet geometry and temperature profile*

*The case of initial temperature $T_i = 25°C$ (ambient)*

First, the temporal variation of droplet geometry and the temperature for the case of initial temperature ($T_i$) = 25°C, i.e., when there is no elevation of droplet initial temperature, will be presented. The variation of droplet geometry with time is shown in figure 4. In Figure 4 (a), the side view of the droplet shapes at different times during evaporation, extracted from high-speed video frames, have been shown. The topmost (0s) image in Figure 4 (a) shows the definitions of the apparent contact angle ($\theta$) and the wetted diameter ($D_W$). Since the droplets are smaller than the capillary number, they are spherical cap in shape. The reflection of the droplet onto the substrate is visible which enables one to detect the triple-phase contact line (contact points) with reasonable accuracy. When there is no elevation of $T_i$, the initial $\theta$ and $D_W$ is 30°±2° and 2.39 mm respectively. The error in $D_W$ is negligibly small compared to its absolute value and hence is not presented in Figure 4 (b). Measurement of $\theta$ is consistent with the contact angle of water on pure glass [46]. From Figure 4 (b), it is seen that during evaporation, $D_W$ remains nearly constant for most of the evaporation time (almost ~ 90 % of the total evaporation time), while the global trend of $\theta$ exhibits a monotonic decrease. Hence, the evaporation happens in CCR mode and the triple-phase contact line remains pinned for most of the evaporation time. At the last stage of evaporation, there is a sharp decrease in $D_W$, indicating a receding of the triple phase contact line. Henceforth, the stage at which the triple-phase contact line starts receding will be termed as the last stage of evaporation.

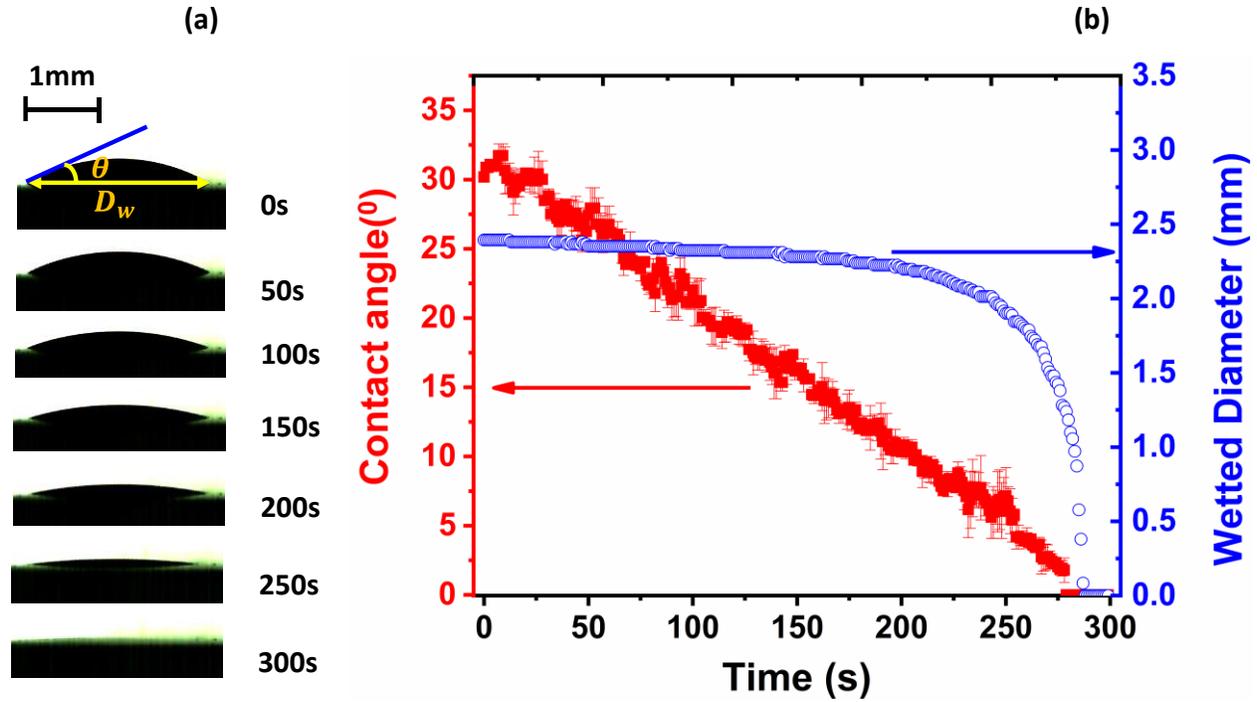

**Figure 4.** Temporal variation of droplet geometry for $T_i=25°C$ (a) high-speed side visualization of the droplet at different evaporation times (b) Temporal variation of contact angle ($\theta$) and wetted diameter ($D_W$) with time, during the evaporation.

The temporal variation for interface temperature (liquid-vapor, solid-vapor) for the case of $T_i= 25°C$ is shown in Figure 5. The IR thermography is also presented in supplementary video V1. In Figure 5, IR images extracted from the time frames of supplementary video V1 is also shown. It is seen that during the whole evaporation process, the droplet apex is relatively cooler than the contact line region which is expected as the apex has a longer thermal conduction path across the droplet than that of the contact line region. A weak temperature gradient (~ 3°) exists across the liquid-vapor interface which is commonly found for a water droplet resting on the glass surface at ambient [34].

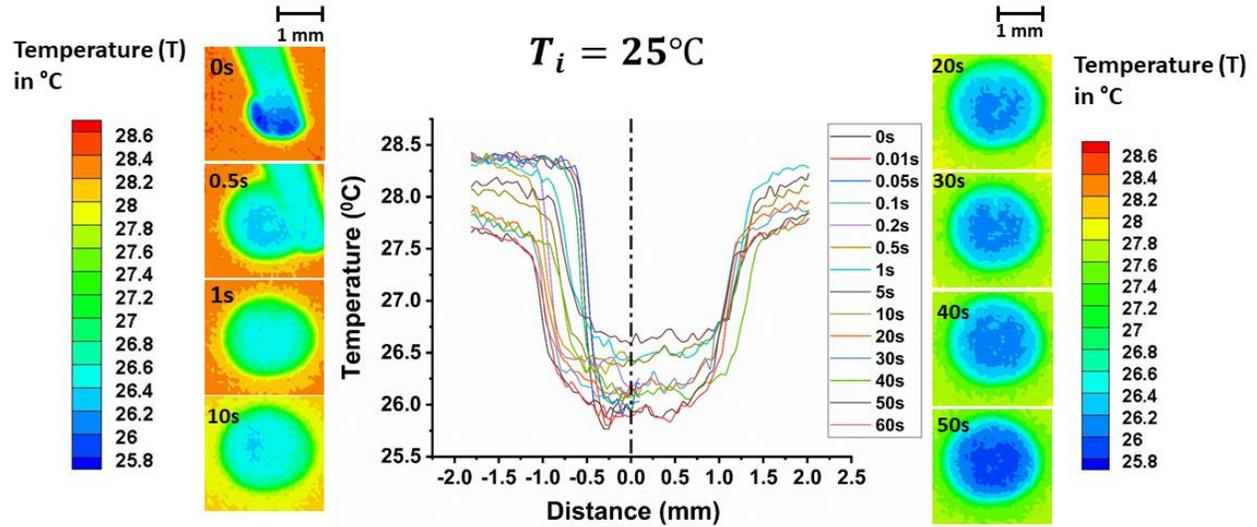

**Figure 5**. Temporal variation of interface temperature for $T_i$=25˚C, representative IR camera images at different times during the evaporation are also shown. See also associated supplementary movie V1, wherein the line along which the plots in the middle column have been obtained is marked.

*The cases of initially heated droplets, $T_i$ = 55˚C and 75˚C*

Figure 6 shows the temporal variation in droplet geometry for the case of elevated initial droplet temperatures- (a) $T_i$=55°C and (b) $T_i$= 75˚C. Similar to the case of $T_i$ = 25˚C, it is found that the evaporation happens mostly in CCR mode except at the last stage of evaporation when there is a rapid drop in the wetted diameter signifying a receding triple-phase contact line. The total evaporation time is slightly lesser (~ 50 s) than that for the case $T_i$ = 25˚C. From these observations, one can say that upon elevation of droplet initial temperature, the total evaporation time does not reduce significantly, and hence, the elevated temperature does not persist for a sufficiently long time. This is further confirmed by IR thermography as discussed next.

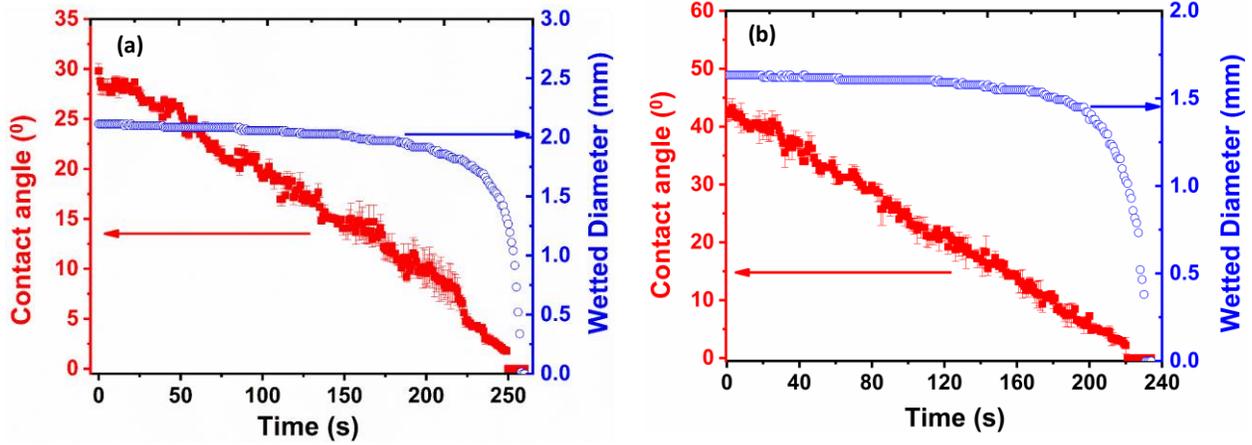

**Figure 6**. Variation of droplet geometry with time during evaporation for the two cases of initial droplet temperature: (a) $T_i = 55°C$ and (b) ) $T_i = 75°C$

Figures 7 and 8 show the temporal variation of interface temperature during evaporation for the cases of $T_i = 55°C$ and $T_i = 75°C$ respectively. The temporal variation of the interface temperature field can also be found in supplementary movies V2-V5. The IR images extracted from time frames of supplementary videos V2-V5 are shown in figures 7 and 8 which depicts the interface temperature field at different instances of time. For the sake of contrast and clarity, the data has been bifurcated into two parts: one from initial 0s-10s (supplementary movies V2, V4 for $T_i = 55°C$ and $T_i = 75°C$ respectively), and another for the rest of the time (supplementary movies V3, V5 for $T_i = 55°C$ and $T_i = 75°C$ respectively). Also, the plots and the IR images have been bifurcated into two-time regimes, as in figures 7 and 8, for better contrast and clarity. Time *t=0* is defined when the droplet comes in contact with the substrate surface and a gradient in the temperature across the liquid-vapor interface is established. It is seen that at the very early stage of evaporation, a significant temperature gradient exists across the liquid-vapor interface- the difference between the droplet apex and the contact line region temperature being $\Delta T = T_{apex} - T_{edge} \sim 25°C$ for $T_i = 55°C$ and $\sim 42°C$ for the case of $T_i = 75°C$. However, this gradient decays rapidly and vanishes within $\sim$ 1s after the deposition of the droplet. After $\sim$ 10s, the system returns to the same state as that of $T_i = 25°C$, cf. Figure 5. This is also depicted in Figure 9, wherein the variation of $\Delta T$ with respect to time is plotted. It is seen that $\Delta T$ have higher values at the early stages of evaporation which decays fast and after $\sim$ 10s, the curves for $T_i = 55°C$ and $T_i = 75°C$ merge with that for $T_i = 25°C$.

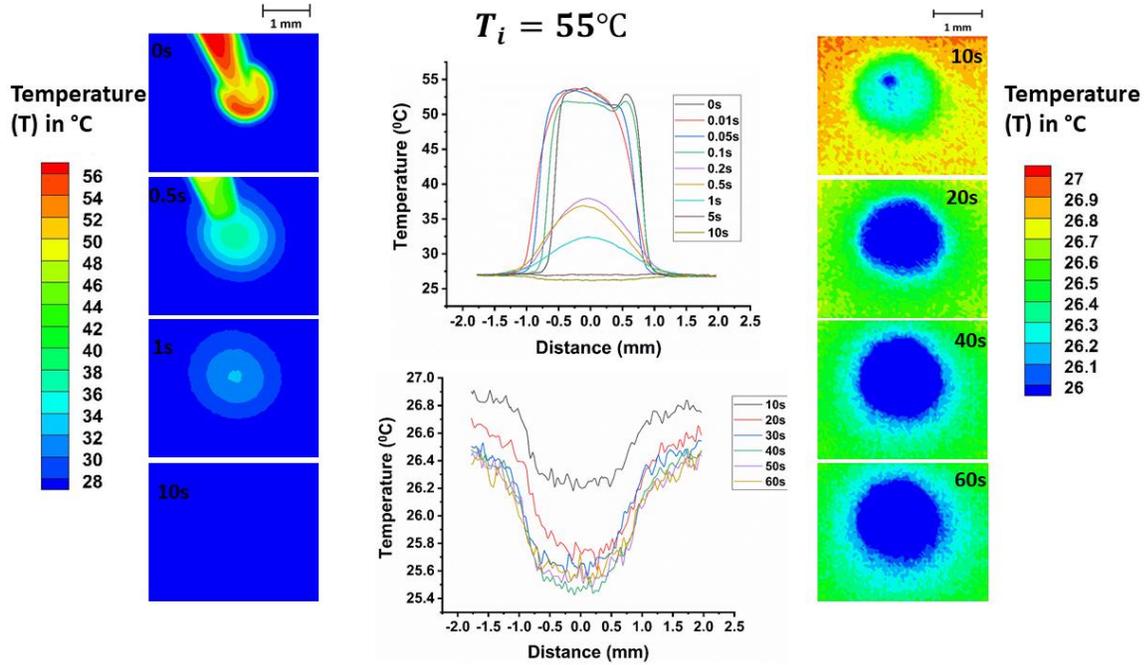

**Figure 7**. Temporal variation of interface temperature during evaporation for the case of $T_i = 55°C$. Representative IR camera images at different times during evaporation are also shown. See also associated supplementary movie V2 and V3, wherein the lines along which the plots in the middle column are obtained, are marked The movies are slowed down ten times for better contrast.

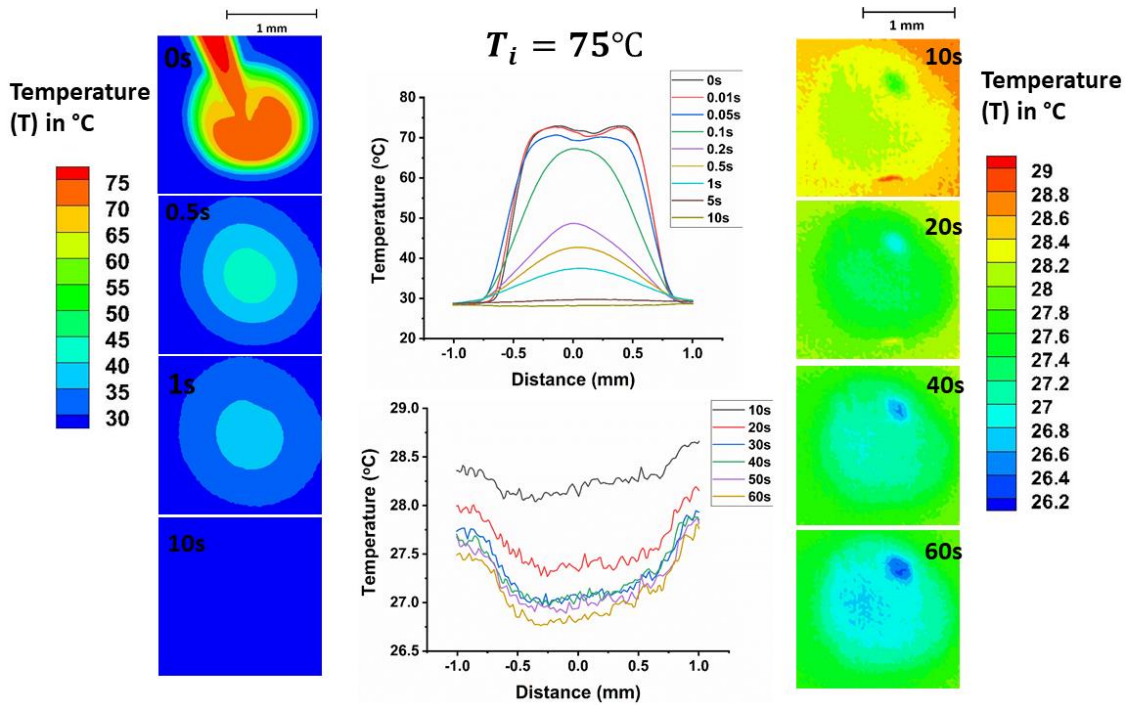

**Figure 8**. Temporal variation of the interface temperature during the evaporation for the case of $T_i = 75°C$. Representative IR camera images at different times during evaporation are also shown. See also supplementary movie V4 and V5, wherein the lines along which the plots in the middle column are obtained, are marked The movies are slowed down ten times for better contrast.

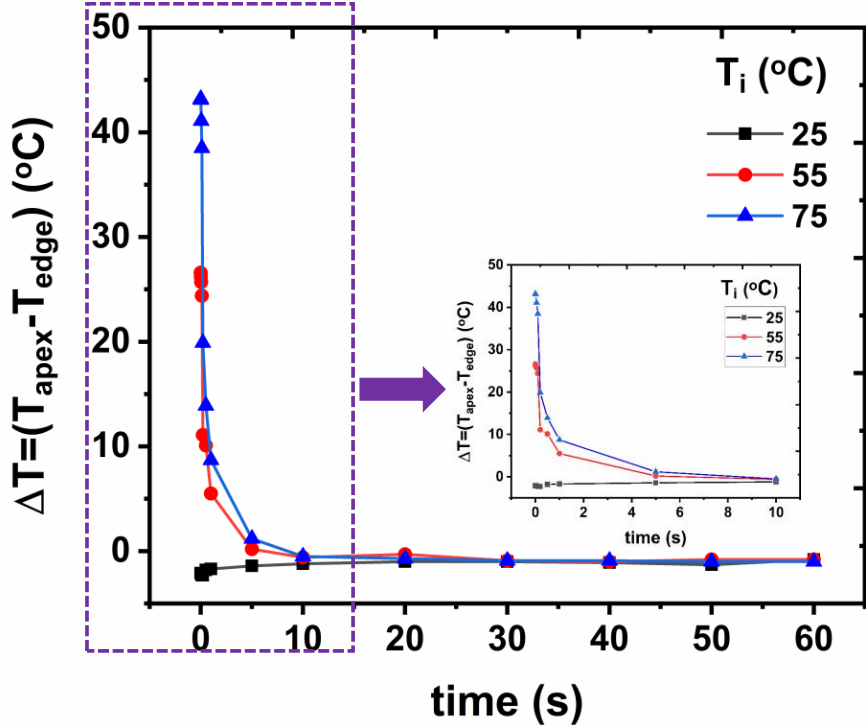

**Figure 9**. The decay of temperature difference $\Delta T$ between droplet apex and edge (contact line region) with time for the cases of $T_i = 25°C$, $T_i = 55°C$ and $T_i = 75°C$. Inset shows the variation of $\Delta T$ for the first 10s for better contrast.

*(ii) Deposition patterns and optical profilometer measurements*

In this section, the resultant dried colloidal deposits due to the elevation of droplet initial temperature will be presented. Representative deposition patterns after evaporation of colloidal sessile droplets under the condition of different initial temperatures are shown in Figure 10 (a). The experiments were repeated at least five times to ensure repeatability. For the case of $T_i = 25°C$, a conventional coffee ring pattern is obtained. We have seen that in the case of $T_i = 25°C$, the temperature gradient along the liquid-vapor interface is small (~ 3°C) in this case. When water is chosen as probe liquid, it was earlier demonstrated [47] that this temperature gradient is insufficient to generate a thermal Marangoni flow. Thus, only radially outward flow is present which leads to a formation of the classical coffee ring pattern for a pinned triple-phase contact line. This implication is consistent with the observation of ring pattern in figure 10(a) for the case of $T_i = 25°C$. The pattern is also consistent with previous observations for the given particle size and concentration [48]. Interestingly, when the droplet initial temperature is elevated to 55°C and 75°C, a significant inner deposit is observed, even though the initial liquid-vapor interface temperature gradient vanishes rapidly within a first few seconds and the condition for $T_i = 25°C$ is obtained for the rest of the evaporation process (cf. figures 7, 8 and 9). This fact is further examined by optical surface profilometry of the outer most ring, representative results of which are shown in Figure 10 (b). The profiling length is shown in the column of Figure 10 (c). Such profilometer data has been acquired in four azimuthal locations on different samples and the average values of ring height and width are reported (cf. Figure 13), the experimental spread being shown by error bars (cf. Figure 13). The procedure for obtaining the average ring profile and dimensions can be found in earlier communications [34, 48], however, it has been further elaborated in the supporting information. It is found that for the case of $T_i = 25°C$, almost three monolayer deposits are found on the ring, whereas, with the elevation of droplet initial temperature, only a single monolayer ring with a diminished ring width is formed. Hence, for the conditions of $T_i = 55°C$ and 75°C, we see a sufficient loss of ring mass (cf. figures 10 and 13) and associated formation of inner deposits.

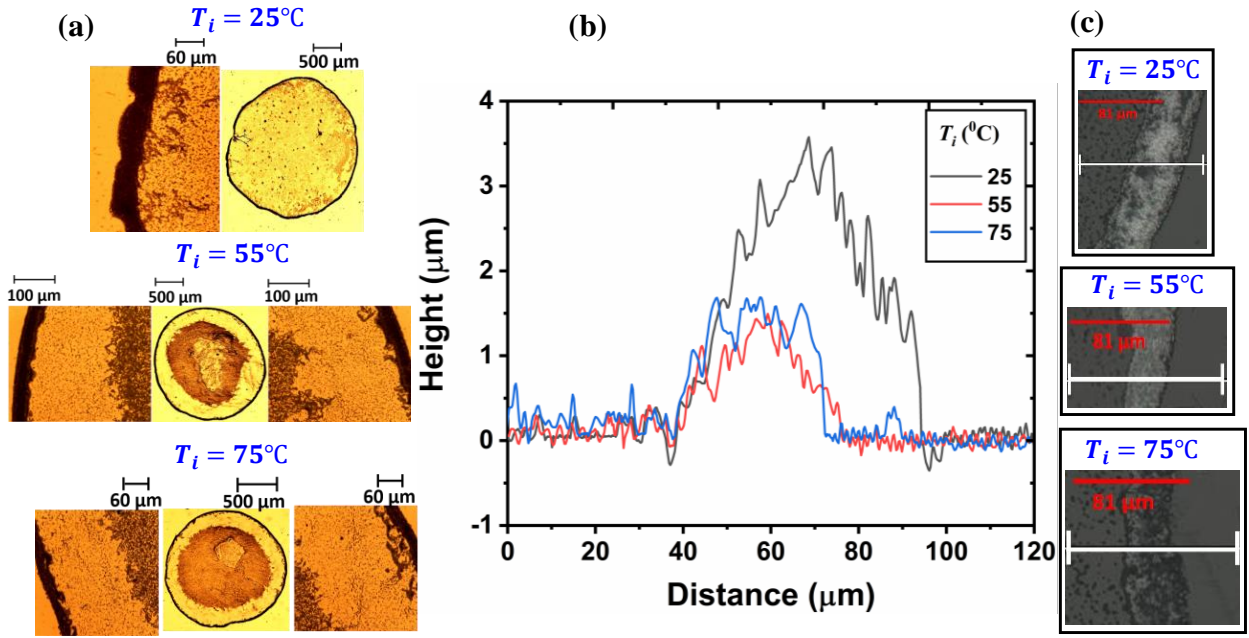

**Figure 10**. Modification of dried colloidal deposit pattern upon elevation of droplet initial temperature (a) representative optical microscopic view of the dried deposit patterns (b) representative ring profiles obtained from optical profilometry (c) surface profiling lengths for (b)

**Effect of particle concentration**

Here the deposition patterns for particle concentrations of $c'$ = 0.05% v/v and 1.0% v/v will be presented. The particle size remains the same- 1.1 μm mean diameter. Representative results are shown in figures 11 and 12 respectively for $c'$ = 0.05% v/v and $c$ = 1.0% v/v, wherein, columns (a) show the microscopic images of the deposit patterns under different initial droplet temperatures, columns (b) show the optical profilometry data for the ring and columns (c) show the profiling lengths for columns (b). Similar to the observations presented for particle concentration $c'$ = 0.1% v/v, here also, it is seen that with elevation in droplet initial temperature, the outer ring diminishes and a significant fraction of the colloidal particles are deposited as inner deposits. For the case of $c' = 0.05\%$ v/v, a slight discontinuity in the outer ring is observed for the cases of $T_i = 55°C$ and $T_i = 75°C$. From the zoomed-in microscopic views of the outer rings in figure 11, it can be seen that, for the cases of $T_i = 55°C$ and $T_i = 75°C$, the particles in the ring are more loosely packed as compared that of $T_i = 25°C$, making the outer ring slightly discontinuous for the cases of higher $T_i$. This may be attributed to the loss of particles mass from the ring at low concentration and can be considered as an additional signature of the fact that the suspended particles are getting advected away from the ring at higher $T_i$. Figure 12 shows the resultant dried deposit pattern for the largest particle concentration ($c' = 1.0\%$ v/v) considered in the present investigation. Notably, the effect of elevation of initial droplet temperature is particularly prominent at higher particle concentration, e.g., $c' = 1.0\%$ v/v as can be seen from Figure 12 (also from Figure 13). From column (b) of figure 12, it can be seen that for the cases of $c' = 1.0\%$ v/v, a sharper reduction in the ring dimensions is recorded as compared to lower particle concentrations, i.e., the thinning of the outer ring with increasing initial droplet temperature is particularly prominent for $c' = 1.0\%$ v/v. In fact, by comparing figures 10, 11, and 12; the effect of elevation of initial droplet temperature becomes increasingly prominent as one progressively goes from lower particle concentration towards higher particle concentration- at higher particle concentrations, more prominent loss of ring mass, and associated inner deposit are found. This is attributed to the following fact. At higher particle concentration, a higher number of particles are available for the advection by the fluid flow, and thereby, a more prominent modification occurs in the dried deposit pattern.

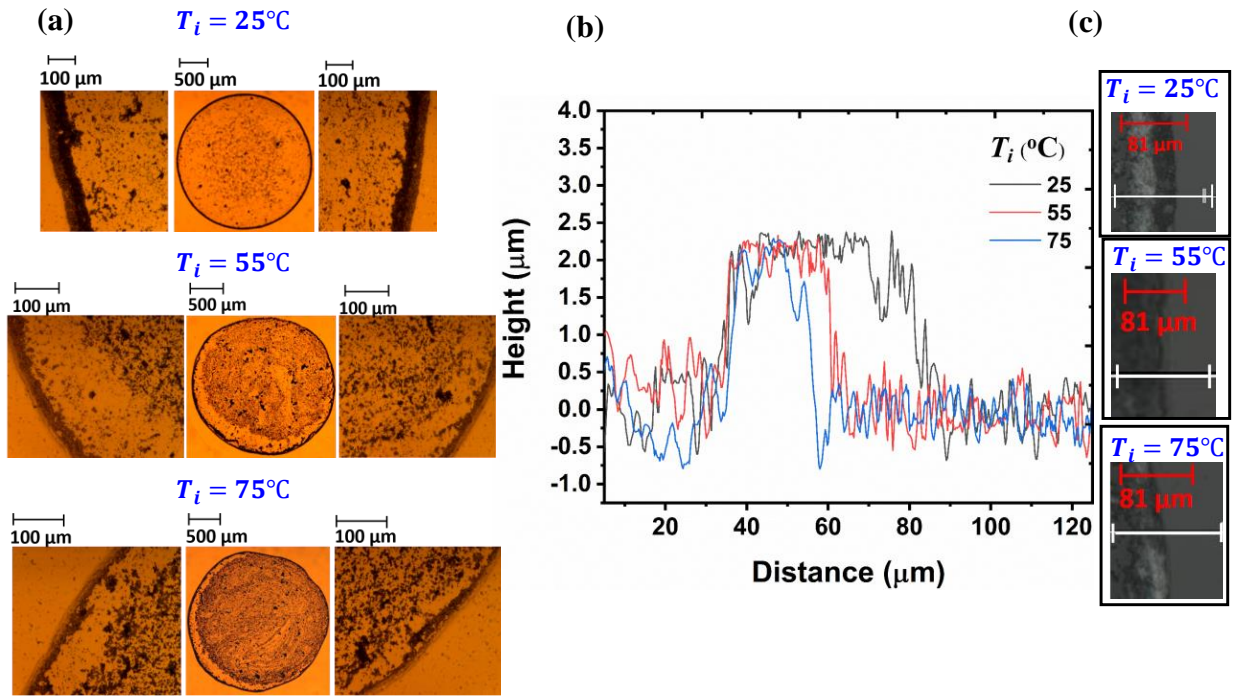

**Figure 11**. Dried colloidal deposit patterns upon elevation of droplet initial temperature for $c' = 0.05\%$ v/v (a) representative optical microscopic view of the dried deposit patterns (b) representative ring profiles obtained from optical profilometry (c) surface profiling lengths for (b)

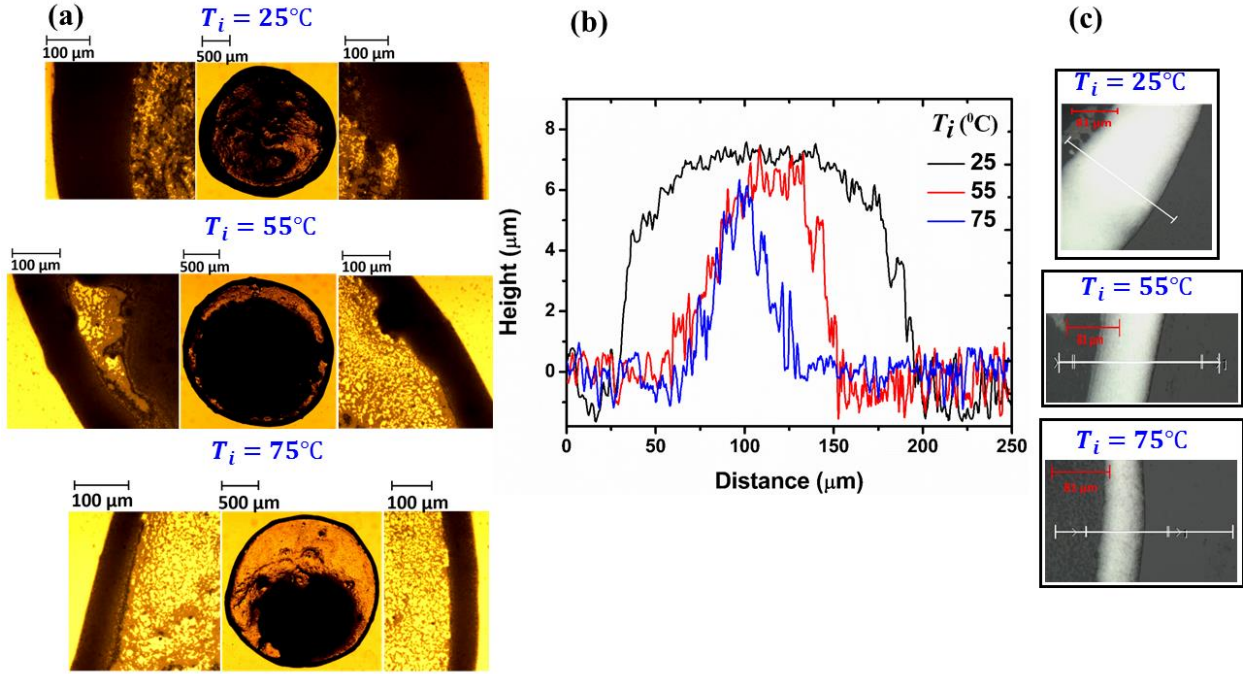

**Figure 12**. Dried colloidal deposit patterns upon elevation of droplet initial temperature for $c' = 1.0\%$ v/v (a) representative optical microscopic view of the dried deposit patterns (b) representative ring profiles obtained from optical profilometry (c) surface profiling lengths for (b)

Figure 13 presents a quantitative comparison of average ring height and ring width with increasing $T_i$ for the three different particle concentrations ($c'$) under consideration. Consistent with the previous discussion, it is clearly seen that with an increase in $T_i$, the outer ring diminishes to a significant extend. We observe a reduction in both average ring height and width with increasing $T_i$ for all the cases of particle concentrations under consideration. Thus, by increasing the initial droplet temperature, a significant loss of particle mass from the ring and associated inner deposits are obtained. This means that as the droplet is elevated to high initial temperatures, the suspended particles move away from the contact line region leading to lesser deposits on the ring and formation of associated inner deposits. Also, the reduction in the ring dimensions with the elevation of initial droplet temperature becomes more and more prominent at higher particle concentrations, providing more experimental contrast. This was also depicted in figures 10-12. Therefore, the present investigation discloses a distinct way for suppression of the coffee-ring effect which may be proven to be useful in industrial processes detailed in the introduction section. In the following sections, we will provide further insights into the mechanism of how the coffee-ring is suppressed due to the elevation of initial droplet temperature.

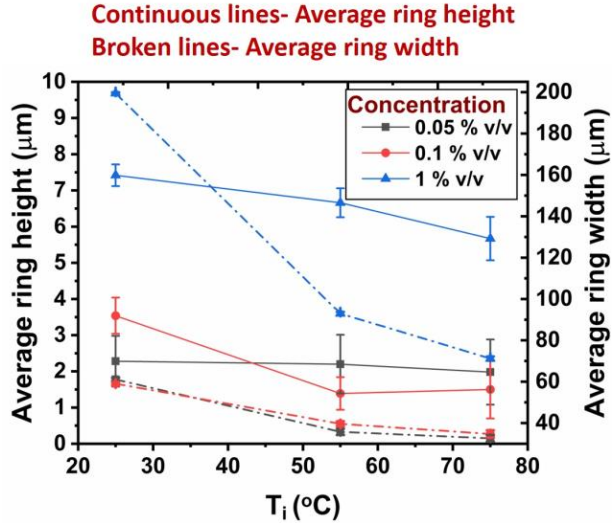

**Figure 13**. Comparison of variation of average ring height and average ring width increasing droplet initial temperature ($T_i$) for different particle concentrations $c' = 0.05\%$ v/v, 0.1% v/v and 1.0% v/v.

**Internal fluid flow visualization by fluorescent colloidal particles**

From the discussion presented so far, it is seen that when an initially heated sessile droplet is deposited on a substrate surface kept at ambient, initially a strong gradient in temperature exists along the liquid-vapor interface. However, this temperature gradient decays very rapidly and vanishes within the first ~ 10s of the evaporation process. Despite this fact, a significant modification in the dried colloidal deposits is found- the dried deposit pattern exhibits an inner deposit with a diminished ring (cf. figures 10-13). It is envisioned that the initial temperature gradient along the liquid-vapor interface establishes strong Marangoni stress giving rise to an axisymmetric vortex-like recirculating flow pattern, as shown schematically in the inset of Figure 14. Earlier, it was demonstrated that depending upon the substrate to liquid thermal conductivity ratio, the direction of the liquid-vapor interfacial temperature gradient would determine the direction of Marangoni recirculation [49]. In the present case, at the early stages of evaporation, the liquid-vapor interface temperature decreases from droplet apex towards the contact line region. Hence, the direction of circulation must be reversed as compared to the case of substrate heating [33-35]. Furthermore, it is hypothesized that the recirculating flow generated by the initial liquid-vapor temperature gradient induced Marangoni stress, continues till the last stage of evaporation as a consequence of liquid inertia and is subject to viscous diffusion. The suspended colloidal particles get advected continuously by this recirculating flow. This results in a loss of mass from the outer ring, and associated inner deposit. The temperature gradient at the later stage is weak (~ 3˚C, cf. figure 5) and is insufficient to generate a Marangoni flow in the opposite direction. The hypothesis made from the experimental observations has been evidenced by observing the internal fluid flow during the whole evaporation process which is discussed below.

To understand the exact nature of the internal fluid flow inside the droplet under the condition of initial elevated temperature, the motion of fluorescent colloidal particles suspended in the droplet was monitored under the optical microscope. Due to experimental constraints, the videos were recorded at t ~50s onwards, after the deposition of the droplet. First, the fluorescent colloidal particle-laden, heated droplet was dispensed on the substrate surface, and thereafter, the substrate containing the droplet was gently inserted under the 10X objective lens of the optical microscope wherein, the exciting ultraviolet radiation was already switched on. Supplementary movies V6 and V7 show the motion of fluorescent particles inside the evaporating droplets having initial temperatures of 25°C and 75°C, respectively. In movie V6, where there is no elevation in the droplet initial temperature, the classical radially outward flow is observed. It is noted that the thickness of the "coffee ring" and the amount of particles in the ring increases gradually with time as evaporation goes on. The direction of the flow further confirms that no additional modification in the evaporation dynamics is induced due to ultra-violet ray irradiation. Supplementary movie V7 represents the fluid motion under the condition of $T_i = 75°C$. Here, the particles exhibit recirculating motion. For the sake of demonstration of flow direction, two representative particles have been chosen and their motion at different instances of time is shown in Figure 14. The different time frames of Figure 14 are extracted from the supplementary movie V7. In figure 14, two oppositely moving particles have been chosen and are marked by white and yellow circles, respectively. Their directions of motion are represented by white and yellow arrows, respectively. The two oppositely moving particles, as depicted in time frames 50 s - 65 s in Figure 14, conclusively show that the fluid flow is a recirculating flow. Further, tracking the motion of the white circled particle, one can see that initially it moves towards the contact line and at a time frame of ~ 80 s, it reverses its motion. It is first entrained along the droplet surface (or liquid-vapor interface) and moves towards the contact line before reversing its direction of motion. Thus, the direction of recirculating motion, as is envisioned in the inset of Figure 14, is evidenced.

The motion is explained as follows. At the initial stage, the droplet is hottest at the apex and is the coldest at the edge, inducing a Marangoni stress along the liquid-vapor interface. This initial stress starts a flow along the liquid-vapor interface which is directed towards the contact line. Therefore, the excess amount of fluid that is transported towards the contact line, has to flow back from the contact line due to mass conservation. This reversal in flow preferentially occurs away from the liquid-vapor interface (nearer to the solid-liquid interface) to minimize the frictional losses with the opposite flow near the liquid-vapor interface [31]. The resultant flow is therefore convective and thus an axisymmetric vortex flow pattern in the clockwise direction is established within the droplet, as shown schematically at the inset of figure 14. It is to be noted that the supplementary movie V7 (or the frames of Figure 14) records the top view of the droplet, and hence the vertical component of particle motion is not revealed. Overall, observation of a bulk flow away from the contact line in conjunction with the reversal of direction of motion of the suspended particles near the contact line is a clear indication of the recirculating flow scenario. The microscope focus was slowly shifted to the bottom (away from the liquid-vapor interface) of the droplet towards the end of evaporation, to keep track of the inward moving particles. Visual observation also reveals that the development of "coffee ring", in supplementary movie V7 ($T_i = 75°C$), is not as prominent as compared to that of supplementary movie V6 ($T_i = 25°C$). Lastly, the recirculating motion continues till the last stage of evaporation, i.e., the stage when the triple-phase contact line starts receding. As it has been shown in the last time frame of Figure 14 (170s), the particles get advected by the recirculating motion and are collected at the middle of the solid-liquid interfacial area before the triple-phase contact line starts receding. Hence, it is evident that the recirculating motion, generated by the initial "impulsive" liquid-vapor interface temperature gradient induced Marangoni stress, continues due to liquid inertia and should be subjected to decay by viscous diffusion mechanism. This notion will be analyzed later with the help of qualitative scaling arguments obtained by the first approximation (cf. figure 19).

In connection with Figure 14, the final deposition pattern for the fluorescent particles under the condition of $T_i = 25°C$ and $T_i = 75°C$ has been shown in Figure 15. While conventional ring pattern is obtained for $T_i = 25°C$, significant inner deposit with the diminished ring is found for $T_i = 75°C$ which is consistent with the observations of Figure 10 (a), given the same particle size and concentration. Thus, by monitoring the motion of fluorescent particle tracker, the origin of inner deposits may be understood.

We also observe the pinning of the contact line (CCR mode) for almost the whole duration of the evaporation period for the case of initially heated droplets. This is also evident from the resultant dried deposition pattern wherein; an outer ring is formed along with the inner deposits. This can be explained by the flow patterns depicted schematically in the inset of Figure 14. After the diffusion of heat (~ 10s after droplet deposition on the substrate), the contact line region remains the hottest part of the droplet and therefore, evaporation driven replenishment flow towards the contact line starts. On the other hand, the initial Marangoni stress-induced recirculating flow continues as is seen in Figure 14. Near the contact line, these two oppositely directed flows and a stagnation region are present as shown schematically in the inset of Figure 14. The similar flow pattern and stagnation region near the contact line were reported earlier [30] for the clockwise direction of Marangoni vortex circulation inside a surfactant laden sessile droplet. The flow pattern in the present case is different from the heated substrate case, where, the liquid-vapor interfacial temperature gradient is in the opposite direction and consequently the Marangoni vortex circulation is in anti-clockwise direction [34]. Also from supplementary movie V7, it is seen that the particles first move towards the contact line and reverse their direction of motion in the region slightly away from the contact line. A few particles are trapped in this stagnation region and lead to the pinning of the contact line, bringing about the formation of the outer ring.

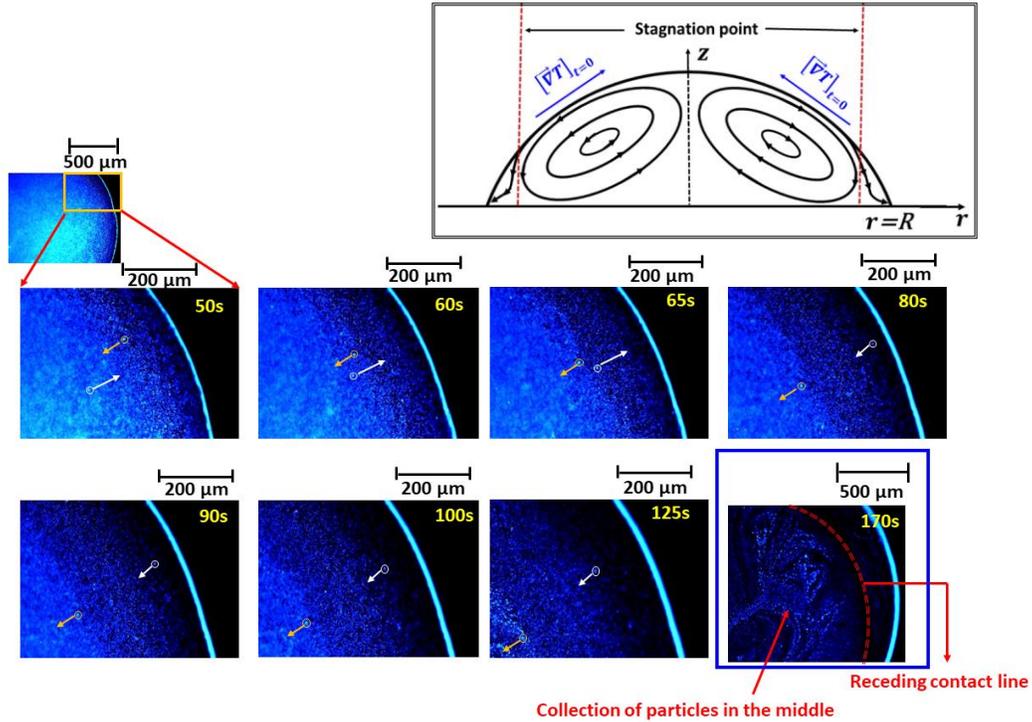

**Figure 14.** The motion of fluorescent particles under the condition of $T_i = 75°C$, two representative particles are shown by white and yellow circles, and their directions of motion are shown by white and yellow arrows respectively. The last frame shows collection of particles in the middle and receding of the triple-phase contact line at the last stage of evaporation, see also associated supplementary movie V7. Inset shows a schematic of the recirculating flow patterns along with the stagnation region.

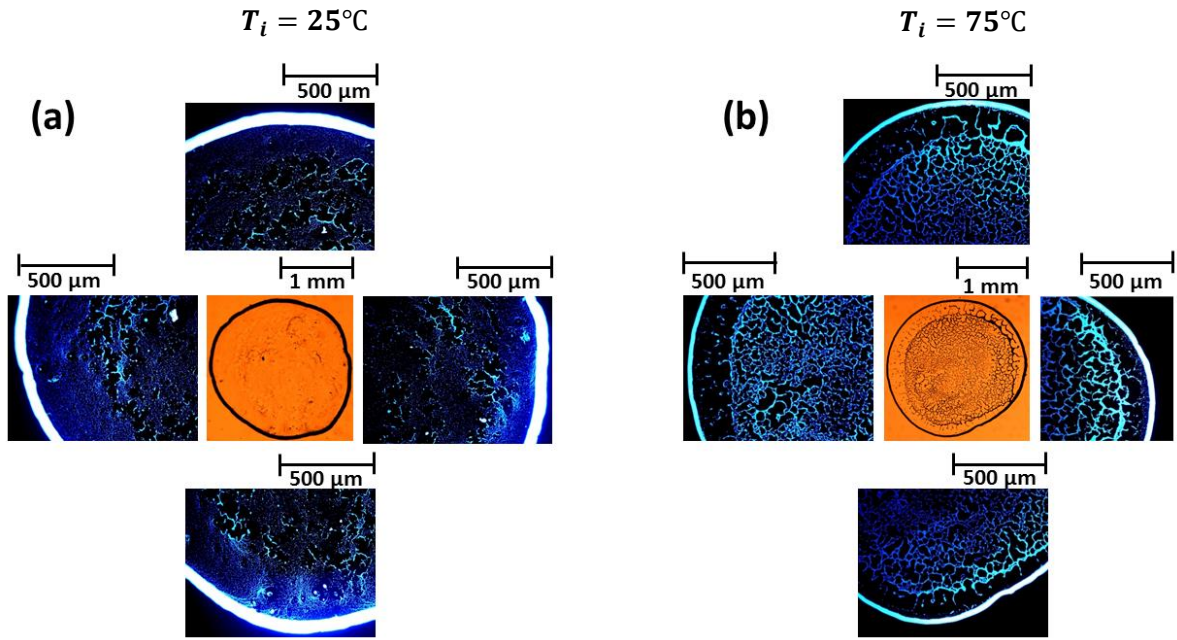

**Figure 15.** Final dried deposit pattern of fluorescent particles after evaporation for initial temperatures ($T_i$) of 25°C and 75°C, the zoomed-in view of the four parts of the ring are also shown.

**Heat and mass transfer: comparison of measurements with the computational model**

In this section, the results of the two-dimensional computational model, described in the methods section, and the comparisons between the numerical and experimental results are presented. Figure 16 shows the time-varying temperature field computed from the model. A computer animation presenting the temporal variation of computed temperature distribution in axisymmetric coordinates can be found in supplementary movies V8 and V9. For the sake of clarity and contrast, the animation has been bifurcated into two-time regimes: one from 0-10s (supplementary movie V8), and the other for the rest of the time (supplementary movie V9). The insets in Figure 16 show the contours of computed isotherms (extracted from the different time frames of supplementary movies V8 and V9) depicting the temperature distribution at different instances of time during evaporation. The temperature along the liquid-gas and solid-gas interfaces is plotted in the middle column of Figure 16 for two-time regimes- 0-10s and onwards.

The measured time-varying temperature distribution (supplementary movies V4 and V5) corroborate with computed temperature results along the liquid-vapor interface (supplementary movies V8 and V9). A comparison between Figure 8 and Figure 16 shows that the computed results match the experimental results with reasonable accuracy- the maximum deviation between model prediction and measurements being ~5-7˚C. The model also predicts a fast decay in the temperature gradient, which is consistent with the measurements. It can be seen in the top frame of the middle column of Figure 16 that, initially, a strong temperature gradient exists along the liquid-vapor interface- the droplet remains hottest at its apex and the temperature profile exhibits a plateau up to a radial distance of ~ 0.5 mm and thereafter, the temperature reduces towards the droplet edge. However, this gradient in temperature decays rapidly and vanishes within an initial ~ 10 s of the evaporation process. After this, the system recovers the situation as that of ambient ($T_i = 25°C$) - the droplet remains coldest at its apex and the temperature is the highest at the droplet edge (bottom frame of the middle column of Figure 16). The temperature difference ΔT between the droplet apex and the contact line computed by the model is plotted and compared with experimental ΔT data in Figure 17. A reasonable agreement between the experimental results and the model prediction is found. Therefore, the experimental results are modeled with reasonable fidelity, by considering transient heat and mass transfer and diffusion-limited evaporation.

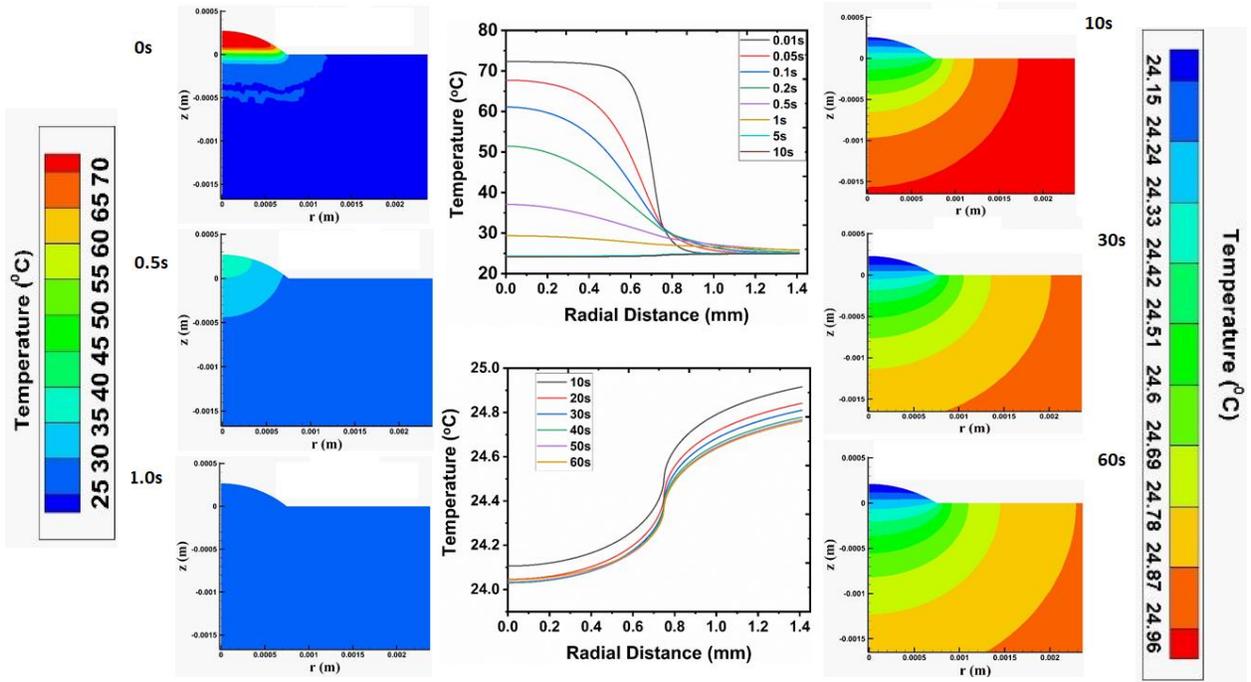

**Figure 16.** Temporal variation of interface temperature during evaporation for the case of $T_i$=75°C, derived from model (for comparison with experimental results see Figure 8). The color images show temperature distribution at different time instances computed from the model (see also associated animations in supplementary movies V8 and V9).

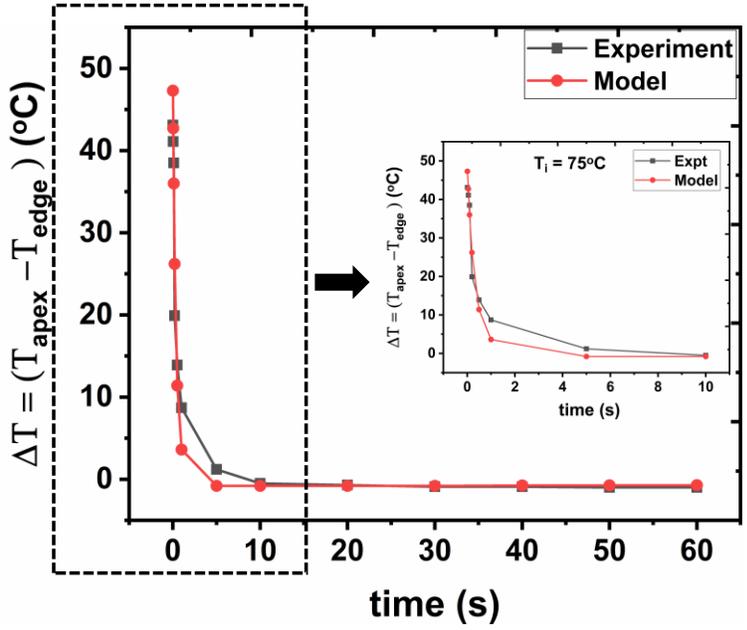

**Figure 17**. Variation of temperature difference between the apex and the edge of the droplet with time- comparison between experiments and model. Inset shows the same for the initial 10s after deposition of the droplet, for better contrast.

The fast decay of the liquid-vapor interfacial temperature gradient can be explained as follows. There are essentially two mechanisms of heat loss (1) thermal conduction through the droplet and the substrate and (2) evaporative cooling of the droplet surface. Also, the thermal diffusivity, $\alpha_j = k_j/(\rho_j\, c_{p,j})$, of glass ($4.4\times10^{-7}$ m$^2$/s) and water ($1.45\times10^{-7}$ m$^2$/s) are on the same order ($10^{-7}$). A combined effect of the thermal diffusion through the droplet and the substrate along with the evaporative cooling causes the temperature gradient to decay fastly. This physics is captured by the model with reasonable fidelity.

The time-variation of evaporation mass flux computed from the model is shown in Figure 18. The animation from the model depicting the temporal variation of the vapor concentration near the droplet surface can be found in supplementary movies V10 and V11. The vapor concentration distribution at different time instances extracted from the time frames of supplementary videos V10 and V11 is also shown in the first and third columns of Figure 18. For the sake of clarity, the animations for the vapor concentration field and the plots of evaporation mass flux are bifurcated into two-time regimes- from 0 s-10 s and onwards. It can be seen that during the initial stages of evaporation, the vapor concentration is larger near the liquid-gas interface ($0 \leq r \leq 0.6$ mm) and the evaporation mass flux (gradient of vapor concentration, eq. 4) is also higher near the droplet apex as compared to the edge (middle column, top frame of Figure 18). The highest evaporation flux is realized at a radial distance of ~ 0.6 mm. This can be understood by looking at the initial temperature profile shown in Figure 16. When the droplet is just deposited on the surface, the top remains the hottest part, which promotes faster evaporation due to large available thermal energy, and the temperature remains almost a constant up to a radial distance of ~ 0.6 mm. Hence, at this distance, the evaporation mass flux becomes the highest due to the higher available surface area as compared to other regions having the same temperature. However, this profile of evaporation mass flux diminishes rapidly in a typical time scale of ~ 10s, which is consistent with the temporal variation of the temperature field. Thereafter, as the temperature profile approaches to that of $T_i = 25°C$, the vapor concentration field and the evaporation mass flux distribution becomes similar to that of ambient ($T_i = 25°C$) as can be seen from figure 18. The evaporation mass flux essentially becomes the highest at the contact line region[36], as can be seen from the lower frame of the middle column of figure 18. The diffusion of the liquid-vapor from the droplet surface towards the far-field is also depicted in the model. Thus, the present model considering diffusion-limited droplet evaporation and involving transient terms in the energy and mass transport equations demonstrates the temporal evolution of the vapor concentration field and the evaporation mass flux for the case of an initially heated droplet drying on a substrate held at ambient air.

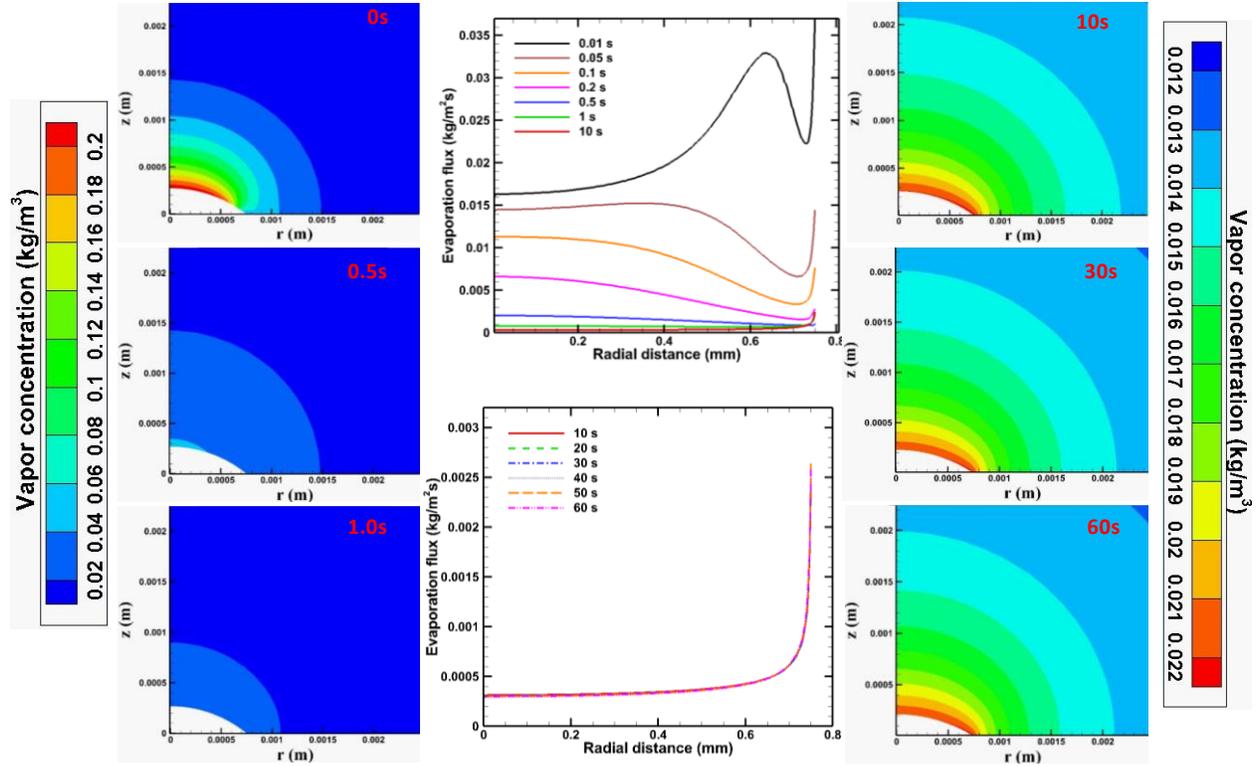

**Figure 18.** Temporal variation of evaporation mass flux during evaporation found from the model for the case of $T_i = 75°C$. Color images show vapor concentration in the vicinity of the droplet surface at different time instances found from the model (see also associated animations in supplementary movies V10 and V11)

## A first approximation treatment of the recirculating flow

It has been shown in previous sections that the temperature gradient along the liquid-vapor interface decays very fast- within the first 10 seconds and thereafter, the system recovers the same situation as that of ambient in which the liquid-vapor interface temperature is the lowest at the droplet apex and remains the highest near the contact line. This implication is also consistent with the computed evaporation mass flux profile shown in Figure 18. Despite this fact, the dried colloid deposit patterns are sufficiently modified for the cases of $T_i = 55°C$, and $T_i = 75°C$ (cf. figures 10-13). It was hypothesized that the initial temperature gradient induced Marangoni stress establishes a flow circulation from the apex to the contact line along the liquid-vapor interface, forming axisymmetric vortex inside the droplet which exists till the end of evaporation, as a consequence of liquid inertia and is subject to viscous diffusion. The temperature gradient at the later stage is weak (~ 3°C, figure 5) and is not sufficient to generate a Marangoni flow in the opposite direction. This hypothesis was further evidenced by recording the motion of suspended fluorescent colloidal particles (cf. figure 14).

Further, a scaling analysis has been performed in an attempt to understand the above-mentioned behavior of the recirculating flow to a first approximation by comparing it with the Lamb-Oseen vortex model [50]. This formulation for viscous diffusion of a radially symmetric, line vortex reads, $u(r',t) = u(r',t=0)\left(1 - e^{-\frac{r'^2}{4\nu t}}\right)$, where $u(r',t)$ is the circulation velocity at a radial distance $r'$ from the vortex core at a time $t$, $\nu$ being the kinematic viscosity (~$0.8 \times 10^{-6}$ m²/s for water at room temperature). In the present analysis, $r'$ has been considered to be one-fourth of the sum of experimentally obtained instantaneous droplet height and wetted radius, which is the typical length scale of the vortex within the droplet and thus the recirculating flow-induced vortex is approximated with a circular vortex. Next, $u(r', t=0)$ has been calculated from the scaling argument that the Marangoni velocity ($v_{ma}$) scales with the temperature difference $\Delta T$ between the apex and the edge of the droplet as [51] $v_{ma} \sim \frac{1}{32}\frac{\beta \theta^2 \Delta T}{\mu}$, where $\beta = d\gamma/dT$ is the gradient of surface tension with respect to temperature (-1.68×10⁻⁴ N/m-K), $\theta$ is the instantaneous contact angle, $\mu$ is the dynamic viscosity (1×10⁻³ N-s/m² for water at ambient), and thus is evaluated from the initial $\Delta T$ found from the experiments (cf. Figure 8). The variation of $u(r',t)$ with $t$ found from this formulation is plotted in Figure 19.

The typical approximated measured fluid flow speeds near liquid-vapor interface are obtained at different times at a fixed radial distance of 0.5 mm (where the flow velocity along the liquid-vapor interface is nearly uniform and the radial velocity component is dominating) from the contact line, by following the motion of the fluorescent particles in supplementary movie V7. In a video frame, corresponding to the time instant at which the speed is to be determined, ten representative particles which are entrained along the liquid-vapor interface exhibiting recirculating motion (since the microscope is focused on the interface), were chosen at the said location by pixel counts (by using ImageJ® software). Thereafter, the distance traveled by the particles in a short time window (2 s) was determined by tracking their locations (by pixel counts) at previous times from the corresponding video frames. Thus, an average particle (fluid flow) speed at that time instant over the distance was obtained. In this approach, the vertical component of the particle velocity is not revealed because the vertical plane is in focus. The so obtained values provide useful qualitative information about the internal fluid flow speed [33]. The measured speeds are plotted (black curve) with the aforementioned Lamb-Oseen analysis (red curve) for the time window of the supplementary movie V7 in the inset of Figure 19. It is seen that there is a reasonable qualitative agreement between the two trends, with a comparable order of magnitude. Hence, this qualitative scaling analysis speaks about how the initial Marangoni stress-induced recirculating flow speed scales with time and thereby, captures the postulate that the flow is subject to liquid inertia which decays by viscous diffusion.

On the other hand, typical averaged capillary (radially outward) flow speed, obtained from supplementary movie V6 ($T_i = 25°C$), is of the order 1.3-2.3 µm/s (blue points in figure 19). The measurements are consistent with the scaling argument that the evaporation-induced radially outward flow speed scales as[51], $v_{rad} \sim \frac{j_{max}}{\rho_L}$ (~ 2.5 µm/s, in the present case), where $j_{max}$ and $\rho_L$ are the maximum evaporation mass flux near the contact line (cf. figure 18) and the liquid density respectively. This is a reference value for the radially outward flow speed, where there is no Marangoni stress-induced circulation. For the case of $T_i = 25°C$, the radially outward flow aids the formation of the coffee ring. In contrast, for the case of initially heated droplets, initially there is a strong Marangoni stress-induced recirculation which decays by viscous diffusion. The order of magnitude of the recirculating flow speed approaches to that of the radially outward flow speed towards the end of the evaporation process. Yet, the recirculating flow is sustained till the end of the evaporation and it continues to be sufficient to suppress the effect of the radially outward flow. The suspended particles are continuously advected by this recirculating flow leading to a suppression of the coffee-ring effect and associated formation of the inner deposits which is evident from supplementary movie V7. A more realistic model to accurately capture the temporal behavior of the vortex pattern requires an extension of the finite element model scheme developed for solving the velocity field inside a drying droplet under Marangoni stress [52] by including transient terms in the fluid flow equation in a manner similar to what has been adopted for temperature and vapor concentration field in the present work and coupling with the temperature and vapor concentration field with accurate boundary conditions. The development of this model is underway and will be presented in future communication.

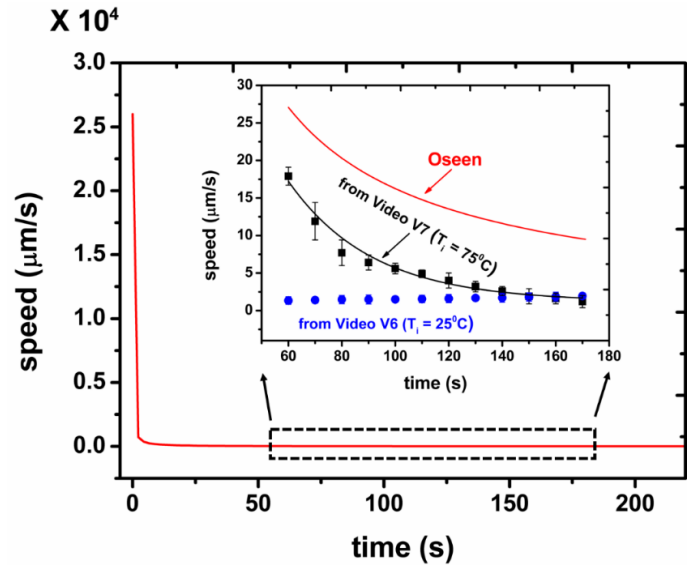

**Figure 19.** Decay of vortex circulation velocity with time found from Oseen model, inset shows a qualitative comparison of the same (red curve) with the speeds obtained at a distance 0.5 mm from the contact line by tracking the fluorescent particles' motion in supplementary movie V7 (black curve). The blue points show the radially outward flow speeds estimated from movie V6. Error bars represent the maximum deviations of the measurements for four recorded movies.

## Conclusions

We have investigated the evaporation and resultant dried colloidal deposit patterns of aqueous sessile droplets having elevated initial temperatures resting on substrates held at ambient air. It was found that an early temperature gradient exists along the liquid-gas interface which triggers a Marangoni recirculation flow from the droplet apex towards the contact line region. Interestingly, the temperature gradient diffuses out rapidly (within initial ~ 10 s after deposition of the droplet) and a condition identical to that of the ambient is recovered for the rest of the evaporation period. Despite this fact, significant suppression in the coffee ring pattern and associated formation of inner deposits are observed. It is envisioned that the initial Marangoni stress-induced recirculating vortex flow continues till the last stage of evaporation as a consequence of liquid inertia leading to significant loss of mass from the outer ring. This hypothesis was further evidenced and confirmed by tracking the motion of suspended fluorescent colloidal particles inside a sessile droplet evaporating under the same experimental conditions. Thus, the suspended colloidal particles are advected by the recirculating flow and the coffee ring is suppressed. A model considering diffusion-limited droplet evaporation and involving transient terms in the energy and mass transport equation corroborates with temporal evolution of temperature profile observed in the experiments and the temporal evolution of evaporative flux has been derived. Finally, the viscous decay of the recirculating internal fluid flow was understood to the first approximation by comparison with a radially symmetric line vortex subjected to viscous diffusion. The present work, therefore, opens a few directions for future work. Firstly, a model for solving the coupled transient heat and mass transfer equations and the fluid flow equations with appropriate boundary conditions can be developed to capture the accurate transient behavior of the initial Marangoni stress-induced vortices. This could be a useful advancement of the present experimental work. Secondly, from an experimental point of view, the idea gained from the present work can be extended to use colloidal particle-laden droplets of liquids having varied viscosity and also by varying ratio of thermal conductivity of substrate to liquid. Several such directions can be taken up for investigations in the future.

## Supporting information

- Evaluation of average ring height and ring width from optical profilometer measurements (Supporting_Info_Chatterjee_et_al)
- Movie V1: Interface temperature distribution for $T_i = 25°C$ found from IR thermography (Temp_25_Expt).
- Movie V2: Interface temperature distribution for $T_i = 55°C$ for initial 10 seconds after deposition of the droplet onto the substrate found from IR thermography (Temp_55_expt_10s).
- Movie V3: Interface temperature distribution for $T_i = 55°C$ for 10 to 60 seconds after deposition of the droplet onto the substrate found from IR thermography (Temp_55_expt_10s-60s).
- Movie V4: Interface temperature distribution for $T_i = 75°C$ for initial 10 seconds after deposition of the droplet onto the substrate found from IR thermography (Temp_75_10s_Expt).
- V5: Interface temperature distribution for $T_i = 55°C$ for 10 to 60 seconds after deposition of the droplet onto the substrate found from IR thermography (Temp_75_10s-60s_Expt).
- Movie V6: Motion of suspended fluorescent particles for $T_i = 25°C$ (Fluorescent_25_degree).
- Movie V7: Motion of suspended fluorescent particles for $T_i = 75°C$ (Fluorescent_75_degree).
- Movie V8: Animation of the temperature distribution of the droplet-substrate system for $T_i = 75°C$, found from the model for the initial 10 seconds of evaporation (Temp_75_10s_MODEL).
- Movie V9: Animation of the temperature distribution of the droplet-substrate system $T_i = 75°C$, found from model for 10-60 seconds of evaporation (Temp_75_10s-60s_MODEL).
- Movie V10: Animation of liquid vapor concentration distribution near the droplet free surface for $T_i = 75°C$, found from model for initial 10 seconds of evaporation (Vap_conc_75_10s_MODEL).
- Movie V11: Animation of liquid vapor concentration distribution near the droplet free surface for $T_i = 75°C$, found from model for 10-60 seconds of evaporation (Vap_conc_75_10s-60s_MODEL).

## Acknowledgments

R.B. gratefully acknowledge financial support by a grant (EMR/2016/006326) from Science and Engineering Research Board (SERB), Department of Science and Technology (DST), New Delhi, India. S. C. gratefully acknowledges institute postdoctoral fellowship from IIT Bombay. The AFM experiments were performed at Industrial Research and Consultancy Center (IRCC), IIT Bombay, and the profilometer measurements were performed in the Metrology Laboratory, Department of Mechanical Engineering, IIT Bombay. The authors thank Prof. Rochish M. Thaokar, Department of Chemical Engineering, IIT Bombay, for Zeta potential measurements. We thank Vishal S. Mahale for providing technical assistance in the experiments.